\documentclass[12pt]{article}
\usepackage{amsmath, amssymb}
\usepackage{graphicx}

\newcommand{\ie}{\textit{i.e.}}
\newcommand{\la}{\langle}
\newcommand{\ra}{\rangle}
\newcommand{\cell}[1]{\makebox[0.4cm]{#1}}

\begin{document}

\title{On the existence of a variational principle for
deterministic cellular automaton models of highway
traffic flow}

\author{Nino Boccara\\
Department of Physics, University of Illinois,\\
Chicago, USA\\
\texttt{boccara@uic.edu}\\
and\\
DRECAM/SPEC, CE Saclay,\\
91191 Gif-sur-Yvette Cedex, France\\
\texttt{nboccara@drecam.saclay.cea.fr}}

\maketitle


\vspace{0.5cm}

{\small{\textbf{Abstract.} It is shown that a variety of
deterministic cellular automaton models of highway traffic
flow obey a variational principle which states that, for a
given car density, the average car flow is a non-decreasing
function of time. This result is established for systems whose
configurations exhibits local jams of a given structure. If
local jams have a different structure, it is shown that either
the variational principle may still apply to systems evolving
according to some particular rules, or it could apply under a
weaker form to systems whose asymptotic average car flow is a
well-defined function of car density. To establish these results
it has been necessary to characterize among all number-conserving
cellular automaton rules which ones may reasonably be considered
as acceptable traffic rules. Various notions such as free-moving
phase, perfect and defective tiles, and local jam play an
important role in the discussion. Many illustrative examples
are given.}}

\section{Introduction} The publication of the Nagel-Schreckenberg
highway traffic flow cellular automaton (CA) model \cite{NaSch}
has attracted much interest. Since then, many papers
describing various CA models of traffic flow have been published
\cite{TT, FI, TTMS, FB, NT1}. Most results concerning the
properties of traffic flow models have been obtained with the help
of either numerical simulations or various extensions of the mean-field
approximation. Only few exact results are known. In the case of the
Fukui-Ishibashi (FI) model \cite{FI}, which is a natural extension of
Wolfram's Rule 184 \cite{W}, Fuk\'s  \cite{F} has recently derived
the exact expression of the average car flow as a function of time.
This result shows that, in the infinite lattice size limit, the
average car flow is a monotonous increasing function of time.
That is, the FI model (and Rule 184 to which it reduces if the speed
limit is equal to 1) obeys a variational principle. The purpose of
this paper is to find out to what extent such a principle remains
valid for other deterministic CA models of traffic flow.

\section{Rules and configurations representations} In CA
models of traffic flow on a circular one-lane highway, the road is
represented by a lattice of $L$ cells with periodic boundary conditions.
Each cell is either empty (in state \texttt{0}) or occupied by a car
(in state \texttt{1}). Since the number of cars traveling a circular
highway is conserved, such a system evolves according to a
number-conserving CA rule. We recently established \cite{BF1} a
necessary and sufficient condition for a one-dimensional $q$-state
($q \geq 2$) deterministic CA rule to be number-conserving, and
studied a few illustrative examples. These rules may be seen as
deterministic evolution rules of one-dimensional closed systems
of interacting particles. Among these rules, some are such that
all particles always move in the same direction. All deterministic
CA models of highway traffic flow are members of this family of rules.

Although most papers on CA models of traffic flow deal with
two-state CA, some number-conserving $q$-state CA rules, with
$q>2$, can also be good candidates \cite{NT2}. In this case, a cell
could, for instance, either represent a longer segment of the highway
capable of accommodating a maximum of $q-1$ cars or the unit
segment of a $(q-1)$-lane highway.

In standard CA modeling it is usually assumed that the configuration
at time $t+1$ is entirely determined by the configuration at time $t$.
Within this restrictive framework, models like \cite{NaSch, TT} are
not standard CA models since the road configuration at time $t+1$
depends on road configurations at time $t$ and $t-1$.

As we shall see, the family of potential models of traffic flow in
terms of deterministic one-dimensional two-state standard cellular
automata is rather rich, and, in this paper, only models of this
type will be considered.

In our discussion of traffic rules, we will not represent
CA rules by their rule tables, but make use of a representation
which clearly exhibits the particle motion. This \textit{motion
representation}, or \textit{velocity rule}, which has been
introduced in \cite{BF2}, may be defined as
follows. If the integer $r$ is the \textit{rule's radius}, list
all the $(2r+1)$-neighborhoods of a given particle represented by
a \texttt{1} located at the central site of the neighborhood. Then, to
each neighborhood, associate an integer $v$ denoting the velocity of
this particle, that is the number of sites this particle will
move in one time step, with the convention that $v$ is positive if
the particle moves to the right and negative if it moves to the left.
For example, Rule 184, defined by
\begin{alignat*}{4}
f_{184}(0,0,0) & = 0, &\quad f_{184}(0,0,1) & = 0,
&\quad f_{184}(0,1,0) & = 0, &\quad f_{184}(0,1,1) & = 1,\\
f_{184}(1,0,0) & = 1, &\quad f_{184}(1,0,1) & = 1,
&\quad f_{184}(1,1,0) & = 0, &\quad f_{184}(1,1,1) & = 1,
\end{alignat*}
is represented by the radius-1 velocity rule:
\begin{equation}
\mathtt{\bullet11}\rightarrow\mathtt{0},\quad
\mathtt{\bullet10}\rightarrow\mathtt{1}.
\label{r184}
\end{equation}
The symbol $\bullet$ represents either a \texttt{0} or a \texttt{1},
\ie, either an empty or occupied site. This representation, which
clearly shows that a car can move to the next-neighboring site on
its right if, and only if, this site is empty, is shorter and more
explicit.

When discussing road configurations evolving according to various
illustrative rules, the knowledge of both car positions and
velocities will prove necessary. Therefore, although we are dealing
with two-state CA rules, we will not represent the state of a cell
by its occupation number (\ie, \texttt{0} or \texttt{1}), but by
a letter in the alphabet $\{\mathtt{e,0,1,\ldots,v_{\max}}\}$
indicating that the cell is either empty (\ie, in state $\mathtt{e}$)
or occupied by a car with a velocity equal to $\mathtt{v}$
(\ie, in state $\mathtt{v}\in\{\mathtt{0,1,\ldots,v_{\max}}\}$).
Note that $\mathtt{v}$ is the velocity with which the car is going to
move at the next time step. Configurations of cells of this type
will be called \textit{velocity configurations} or
\textit{configurations} for short.

\section{The deterministic FI model of traffic flow}
The simplest deterministic CA model of traffic flow is the
FI model \cite{FI} which might be defined as follows.
If $d_i(t)$ is the distance, at time $t$, between car $i$ and
car $i+1$ (cars are moving to the right), velocities are updated
in parallel according to the subrule:
\begin{equation}
v_i(t+1) = \min(d_i(t)-1, v_{\max})
\label{R1}
\end{equation}
where $v_i(t)$ is the velocity of car $i$ at time $t$; then cars
move according to the subrule:
\begin{equation}
x_i(t+1) = x_i(t) +  v_i(t+1),\label{R2}
\end{equation}
where $x_i(t)$ is the position of car $i$ at time $t$. The model
contains two parameters: the speed limit $v_{\max}$, which is
the same for all cars, and the car density $\rho$.

\begin{figure}[ht]
\begin{center}
\includegraphics[scale=1.2]{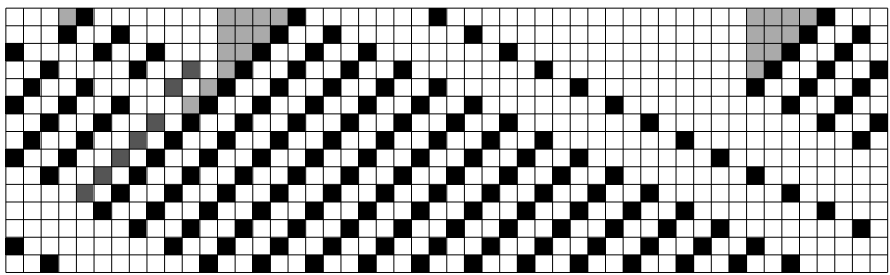}
\caption{\small{Spatiotemporal pattern of rule (\ref{FI2}) for $\rho=0.26$.
States color code: white = $e$, light grey = $0$, dark grey = $1$,
black = $2$.}}
\label{FI2d026}
\end{center}
\end{figure}

\begin{figure}[ht]
\begin{center}
\includegraphics[scale=1.2]{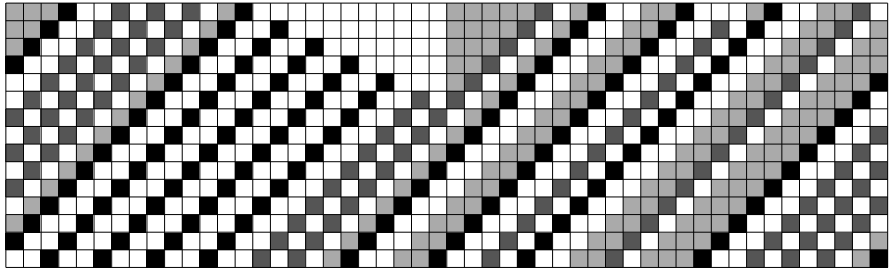}
\caption{\small{Spatiotemporal pattern of rule (\ref{FI2}) for $\rho=0.50$.
States color code: white = $e$, light grey = $0$, dark grey = $1$,
black = $2$.}}
\label{FI2d050}
\end{center}
\end{figure}

For the sake of simplicity, in our discussion of this model, it
is sufficient to consider the case $v_{\max}=2$. The corresponding
radius-2 velocity rule is:
\begin{equation}
\mathtt{\bullet\bullet 11\bullet\rightarrow 0},\quad
\mathtt{\bullet\bullet 101\rightarrow 1},\quad
\mathtt{\bullet\bullet 100\rightarrow 2}.
\label{FI2}
\end{equation}
Figures \ref{FI2d026} and \ref{FI2d050} show examples of spatiotemporal
patterns for car densities $\rho$, respectively, equal to 0.26 and 0.50.
Empty cells are white while cells occupied by a car with velocity $v$
equal to 0, 1, and 2 are, respectively, light grey, dark grey and black.


According to velocity rule (\ref{FI2}), and our choice of configurations
representation, a cell occupied by a car with velocity $v$ must be
preceded by, at least, $v$ empty cells. Each configuration is,
therefore, a concatenation of the following four types of tiles:
\begin{center}
\begin{tabular}{|c|c|c|}
\hline
\cell{\tt 2}&\cell{\tt e}&\cell{\tt e}\\
\hline
\end{tabular}
\hspace{0.7cm}
\begin{tabular}{|c|c|}
\hline
\cell{\tt 1}&\cell{\tt e}\\
\hline
\end{tabular}
\hspace{0.7cm}
\begin{tabular}{|c|}
\hline
\cell{\tt 0}\\
\hline
\end{tabular}
\hspace{0.7cm}
\begin{tabular}{|c|}
\hline
\cell{\tt e}\\
\hline
\end{tabular}
\end{center}
The first tile, which corresponds to cars moving at the speed limit
$v_{\max}=2$, will be called a \textit{perfect tile}, the next two
tiles, corresponding to cars with a velocity less than 2 (here 1
and 0) will be called \textit{defective tiles}, and the last tile
will be called a \textit{free empty cell}, that is, an empty cell
which is not part of either a perfect or defective tile.

Figure \ref{FI2d026} shows that only the first configurations
contain defective tiles. After a few time steps, these tiles
progressively disappear, and the last configurations contain only
perfect tiles and free empty cells. Hence, all cars move at $v_{\max}=2$,
and the system is said to be in the \textit{free-moving phase}. In
Figure \ref{FI2d050}, at the beginning, the same process of
annihilation of defective tiles takes place but, in this case, all
defective tiles do not eventually disappear. A few cars move
at $v_{\max}$, while other cars have either a reduced speed ($v=1$)
or are stopped ($v=0$). This regime is called the \textit{jammed phase}.
For $v_{\max}>2$, FI models exhibit similar qualitative features.

To analyze the annihilation process of defective tiles in FI models,
we need to define what we call a local jam.

\vspace{0.2cm}

\textbf{Definition 1} \textsl{In deterministic FI models of
traffic flow, a local jam is a sequence of defective tiles
preceded by a perfect tile and followed by either a perfect
tile or free empty cells.}

\vspace{0.2cm}

From this definition, it follows that:

\vspace{0.2cm}

\textbf{Proposition 1} \textsl{In the case of deterministic FI
models of traffic flow, the number of cars which belong to a
local jam is a non-increasing function of time.}

\vspace{0.2cm}

This result is a direct consequence of the fact that, by definition,
a local jam is preceded by a car which is free to move, and, according to
whether a new car joins the local jam from behind or not, the number
of cars in the local jam remains unchanged or decreases by one unit.
Note that \textit{the jammed car just behind the free-moving car
leading the local jam becomes itself free to move at the next
time step}.

In order to establish the variational principle, let analyze more
precisely how the structure of the most general local jam
changes in one time step. A local jam consisting of $n$ defective
tiles is represented below:
$$
\cdots v_1\underbrace{ e e \cdots e}_{v_1}
v_2\underbrace{e e \cdots e}_{v_2}
\cdots\cdots v_n\underbrace{e e \cdots e}_{v_n}
v_{\max}\underbrace{e e \cdots e}_{v_{\max}}\cdots,
$$
where, for $i=1,2,\ldots,n$, $0\leq v_i<v_{\max}$. At the next time
step, a car with velocity $v_i$ located in cell $k$ moves to cell
$k+v_i$. Hence, if the local jam is followed by $v_0$ free empty
cells, where $v_0\geq 0$, we have to distinguish two
cases:

(i) If $v_0+v_1<v_{\max}$, then the number of jammed cars remains
unchanged but the leftmost jammed car, whose velocity was $v_1$, is
replaced by a jammed car whose velocity is $v_0+v_1$:
\begin{align*}
\cdots v_{\max}\overbrace{e e \cdots e}^{v_{\max}+ v_0}
v_1 e e \cdots e
v_2 e e \cdots\ e \cdots\cdots & v_n e e \cdots e
v_{\max} e e \cdots e \cdots \\
\cdots e (v_0+v_1) e e \cdots e
v_2 e e \cdots e \cdots\cdots
v_n & e e \cdots e v_{\max} e e \cdots e \cdots
\end{align*}

(ii) If $v_0+v_1\geq v_{\max}$, then the local jam loses its leftmost
jammed car:
\begin{align*}
\cdots e v_1 e e \cdots e
v_2 e e \cdots e \cdots\cdot\cdots & v_n e e \cdots e
v_{\max} e e \cdots e \cdots \\
\cdots v_{\max}\underbrace{e e \cdots e}_{v_{\max}+ v_0'}
 v_2 e e \cdots e \cdots\cdots v_n & e e \cdots e
v_{\max} e e \cdots e \cdots
\end{align*}
and at the next time step, the local jam is followed by
$v_0+v_1-v_{\max} = v_0' < v_0$ free empty cells.

If we partition the lattice in tiles sequences whose end points
are perfect tiles, then, between two consecutive perfect tiles,
either there is a local jam, and the above proof shows that the
number of free empty cells between the two perfect tiles cannot
increase, or there is no local jam, and the number of free empty
cells between the two perfect tiles remains unchanged. We can,
therefore, state:

\vspace{0.2cm}

\textbf{Proposition 2} \textsl{In the case of deterministic FI
models of traffic flow, the number of free empty cells is a
non-increasing function of time.}

\vspace{0.2cm}

\textbf{Remark 1} If a configuration contains no perfect
tiles, then it does not contain free empty cells. Such a
configuration belongs, therefore, to the limit set, and
as we shall see below, the system is in its steady state.
On a circular highway, if a configuration contains only
one perfect tile, the above reasoning applies without
modification.

\textbf{Remark 2} The above proof shows that at each time
step, the rightmost jammed car of a local jam moves one
site to the left. Local jams can only move backwards.

If $L$ is the lattice length, $N$ the number of cars, and
$N_{\rm fec}(t)$ the number of free empty cells at time $t$,
we have
$$
N_{\rm fec}(t) = L-N-\sum_{i=1}^N v_i(t),
$$
since, at time $t$, car $i$ is necessarily preceded by $v_i(t)$
empty cells. Dividing by $L$ we obtain
$$
\frac{N_{\rm fec}(t)}{L}
= 1 - \frac{N}{L} - \frac{N}{L}\,\frac{1}{N}\sum_{i=1}^N v_i(t).
$$
Hence, for all $t$,
\begin{equation}
\rho_{\rm fec}(t) = 1 - \rho - \rho\la v\ra_t,
\label{fec}
\end{equation}
where $\rho_{\rm fec}(t)$ is the density of free empty cells at time
$t$, and $\la v\ra_t$ the average car velocity at time $t$. This
last result shows that, when time increases, since the density of
free empty cells cannot increase, the average car flow
$\rho\la v\ra_t$ cannot decrease. Deterministic FI models of
highway traffic flow obeys, therefore, the following
variational principle:

\vspace{0.2cm}

\textbf{Proposition 3} \textsl{In the case of deterministic FI
models of traffic flow, for a given car density $\rho$, the
average car flow is a non-decreasing function of time and
reaches its maximum value in the steady state.}

\vspace{0.2cm}

The annihilation process of defective tiles stops when there are
either no more defective tiles or no more free empty cells.
Since a perfect tile consists of $v_{\max}+1$ cells, if the car
density $\rho$ is less than $\rho_c=1/(v_{\max}+1)$, there
exist enough free empty cells to annihilate all the defective tiles,
and all cars become eventually free to move. If $\rho>\rho_c$,
there are not enough free empty cells to annihilate all defective
tiles, and eventually some cars are not free to move at $v_{\max}$.
The threshold value $\rho_c$ is called the \textit{critical density}.
At the end of the annihilation process, all subsequent configurations
belong to the limit set, and the system is either said to be
\textit{in equilibrium} or \textit{in the steady state}.

Note that the existence of a free-moving phase for a car density
less than $\rho_c=1/(v_{\max}+1)$, can be seen as a consequence of
Relation (\ref{fec}). When $t$ goes to infinity, according to
whether $\rho_{\rm fec}$ is positive or zero, this relation
implies
$$
\la v\ra_\infty = \min\left(\frac{1-\rho}{\rho}, v_{\max}\right).
$$

If the system is finite, its state becomes eventually periodic
in time, and the period is equal to the lattice size or one
of its submultiples.

Since local jams move backwards and free empty cells
move forwards, equilibrium is reached after a number of
time steps proportional to the lattice size.

\textbf{Remark 3} In the case of Rule 184, the existence of a
free-moving phase for a particle density $\rho$ less than the
critical density $\rho_c=\frac{1}{2}$, obviously implies that the
average velocity $\la v\ra_t$, as a function of time $t$, is
maximum when $t\to\infty$. For $\rho>\frac{1}{2}$, this property
is still true since the dynamics of the holes (empty sites) is
governed by the conjugate of rule Rule 184
(\ie, Rule 226),\footnote{If $f$ is an $n$-input two-state
deterministic CA rule, its conjugate, denoted $Cf$, is defined by
$$
Cf(x_1, x_2, \cdots, x_n) = 1-f(1-x_1, 1-x_2, \cdots, 1-x_n).
$$}
which describes exactly the same dynamics as Rule 184 but for
holes moving to the left. Therefore, for all values of the particle
density, the average velocity takes its maximum value in the
steady state.

In the next section, we shall examine to what extent the
variational principle, valid for all deterministic
FI traffic flow models, remains valid for more general
deterministic CA models of highway traffic flow.

\section{Variational principle} In order to extend Proposition 3
to more general deterministic CA models of traffic flow, we have
first to characterize, among the class of number-conserving
deterministic two-state CA rules, which rules may be considered
as acceptable CA traffic rules.

\subsection{Unidirectional motion}
The first obvious condition to be satisfied by a traffic rule is that
all cars should move in the same direction, that is, in the motion
representation, all velocities should have the same sign.

\textbf{Example 1} For instance,
the rule:
\begin{equation}
\mathtt{\bullet 011\bullet}\rightarrow\mathtt{-1},\quad
\mathtt{\bullet 111\bullet}\rightarrow\mathtt{0},\quad
\mathtt{\bullet\!\bullet\!101}\rightarrow\mathtt{0},\quad
\mathtt{\bullet\!\bullet\!100}\rightarrow\mathtt{1}
\label{r60200}
\end{equation}
is not acceptable since a particle can move either to the right
or to the left. Many number-conserving CA rules are not unidirectional.
In the case of (\ref{r60200}), it can be shown \cite{BF2} that each
particle performs a non-Gaussian pseudo-random walk. The rule being
deterministic, the randomness comes from the randomness of the
initial configuration.

\subsection{Existence of a free-moving phase}
The second natural condition, which should be satisfied by a
deterministic CA traffic rule, is that, when the car density is
sufficiently low, each car should eventually be able to move at
the speed limit $v_{\max}$.

\textbf{Example 2} A numerical simulation of a one-dimensional
system of particles evolving according to the rule:
\begin{equation}
\mathtt{\bullet\!\bullet\!11\bullet\rightarrow 0},\quad
\mathtt{\bullet 1101\rightarrow 1},\quad
\mathtt{\bullet 010\,\bullet\rightarrow 1},\quad
\mathtt{\bullet 1100\rightarrow 2}
\label{R4}
\end{equation}
shows that, for a particle density $\rho<\frac{1}{2}$, all
particles eventually move at $v=1$. For $\rho>\frac{1}{2}$, the
asymptotic average velocity decreases monotonically and goes
to zero for $\rho=1$. This rule is, however, not an acceptable
traffic rule. The velocity of a particle can be equal to $v_{\max}=2$
only if there is another particle located just behind
(\ie, on its left) whose velocity is zero. This implies that,
while $v_{\max}=2$, the average velocity cannot be larger than 1.
The regime in which all particles move at the same velocity $v=1$
is not a true free-moving phase.

\textbf{Example 3} It is often necessary to look closely at the
velocity rule to tell
if a system of particles evolving according to such a rule
exhibits a free-moving phase at low density. Consider, for instance,
the rule:
\begin{align}
&\mathtt{1\!\bullet\!11\bullet\rightarrow 0}, \quad
\mathtt{0111\bullet\rightarrow 0}, \quad
\mathtt{001\!\bullet\!\bullet\rightarrow 0},\notag\\
&\qquad\quad\mathtt{1\!\bullet\!10\,\bullet\rightarrow 1}, \quad
\mathtt{0110\,\bullet\rightarrow 1}.
\label{R17}
\end{align}
The rule elements $\mathtt{1\!\bullet\!10\,\bullet\rightarrow 1}$
and $\mathtt{0110\,\bullet\rightarrow 1}$ indicate
that, if the particle density is too low, no motion can probably
take place. Actually, a simple numerical simulation shows that
below $\rho_{\min}=\frac{1}{3}$, the asymptotic average velocity
is always zero. More precisely, it can be shown that, in the
steady state, if $\rho_v$ denotes the fraction of all particles
whose velocities is equal to $v$, there exist three regimes:

(i) If $\rho\leq\frac{1}{3}$, then $\rho_0 =\rho$.

(ii) If $\frac{1}{3}\leq\rho\leq\frac{1}{2}$, $\rho_0$ and
$\rho_1$ satisfy the relations:
$$
\rho_0 + \rho_1 = \rho,\quad\mbox{and}\quad
3\rho_0 + 2\rho_1 = 1,
$$
that is,
$$
\rho_0 = 1 - 2\rho,\quad\mbox{and}\quad\rho_1 = 3\rho - 1.
$$

(iii) If $\frac{1}{2}\leq\rho\leq 1$, $\rho_0$ and $\rho_1$
satisfy the relations:
$$
\rho_0 + \rho_1 = \rho,\quad\mbox{and}\quad \rho_0 +
2 \rho_1 = 1,
$$
that is,
$$
\rho_0 = 2\rho - 1,\quad\mbox{and}\quad\rho_1 = 1 - \rho.
$$
A system of particles evolving according to (\ref{R17}) does not
exhibit a free-moving phase. It is only when $\rho$ is exactly
equal to $\frac{1}{2}$ that all particles move at $v_{\max}=1$.
As a traffic rule (\ref{R17}) should be discarded.

\textbf{Example 4} The existence of a free-moving phase for a
system of particles evolving according to a given
number-conserving deterministic CA rule does not, however,
guarantee that such a rule is a reasonable traffic rule.
For example, in the case of the velocity rule:
\begin{align}
& \mathtt{\bullet\bullet\bullet 111\,\bullet\rightarrow 0},\quad
\mathtt{\bullet\bullet\bullet 1101\rightarrow 1},\quad
\mathtt{\bullet\bullet\bullet 1011\rightarrow 1},\notag\\
& \mathtt{\bullet\bullet\bullet 1100\rightarrow 2},\quad
\mathtt{\bullet\bullet\bullet\,1010\rightarrow 2},\quad
\mathtt{\bullet\bullet\bullet 100\,\bullet\rightarrow 2},
\label{R7}
\end{align}
all velocities have the same sign, and a simple numerical
simulation shows that, below a critical density equal to
$\frac{1}{2}$, there exists a free-moving phase in which
all the particles move with the velocity $v_{\max}=2$. Its
flow diagram\footnote{Traffic engineers call \textit{fundamental
diagram} the plot of the asymptotic average flow
$\rho\la v\ra_\infty$ versus the density $\rho$.} has even a
nice tent-shape. But, as a traffic rule, (\ref{R7}) shows that
drivers are anticipating the motion of cars on their right \cite{FB}.
Since all drivers move according to the same rule, no collisions
occur. But, to make traffic rules somewhat more realistic, most
authors \cite{NaSch, FI, Schad} consider essential to add some
randomization, as \textit{random braking}, to the basic
deterministic model. In the case of rule (\ref{R7}) this would
lead to collisions, whose number would increase
with the braking probability. Moreover, at the critical density,
systems evolving according to a deterministic car traffic rule
exhibit a second-order phase transition, and it has been
recently shown \cite{BF3} that, for traffic flows evolving
according to FI traffic rules, random braking
is the symmetry-breaking field conjugate to the
order parameter defined as $m = v_{\max} - \la v\ra_\infty$. For
all these reasons, we should not accept as a traffic rule any
deterministic CA rule in which a particle can move to an occupied
site ``knowing'' that the particle located at that site will also move.

In the light of the preceding examples, we shall adopt the following
definition of a free-moving phase

\vspace{0.2cm}

\textbf{Definition 2} \textsl{A system of particles evolving according to
a unidirectional deterministic CA rule exhibits a free-moving phase
if there exists a number $0<\rho_c<1$, called the critical
density, such that, starting from a random configuration with a
particle density $\rho<\rho_c$, the system evolves
to an equilibrium state in which, with probability one, all
configurations consists of perfect tiles and free empty cells.
If the maximum velocity at which a particle can move is $v_{\max}$,
a perfect tile consists of a cell in state $v_{\max}$ preceded by
$v_{\max}$ empty cells.}

\vspace{0.2cm}

\textbf{Example 5} The structure of the last element of the velocity rule:
\begin{equation}
\mathtt{\bullet\!\bullet\!11\bullet \rightarrow 0},\quad
\mathtt{1010\, \bullet \rightarrow 0},\quad
\mathtt{0010\,\bullet \rightarrow 1},\quad
\mathtt{\bullet 1101 \rightarrow 1},\quad
\mathtt{\bullet 1100 \rightarrow 2}
\label{R8}
\end{equation}
shows that the average velocity can never be equal to
$v_{\max}=2$. A system of particles evolving according to such a
rule cannot, therefore, exhibit a free-moving phase in the
sense of the above definition. This type of result is general:
\textit{If a particle can move at $v_{\max}$ if, and only if,
the site located immediately behind it has to be occupied by
another particle, then no free-moving phase in the sense of
Definition 2 can exist}.

\textbf{Example 6} The velocity rule:
\begin{equation}
\mathtt{\bullet\bullet\bullet11\,\bullet\,\bullet\rightarrow 0},\quad
\mathtt{\bullet\bullet\bullet 101\,\bullet\rightarrow 0},\quad
\mathtt{\bullet\bullet\bullet 1001\rightarrow 1},\quad
\mathtt{\bullet\bullet\bullet 1000\rightarrow 2}
\label{OC2}
\end{equation}
describes the behavior of overcautious drivers who avoid occupying
a site located just behind another car. The perfect tile:
\begin{center}
\begin{tabular}{|c|c|c|c|}
\hline
\cell{\tt 2}&\cell{\tt e}&\cell{\tt e}&\cell{\tt e}\\
\hline
\end{tabular}
\end{center}
has not the structure required by Definition 2. It contains an
extra empty cell. While it could be reasonable to consider models
of traffic flows in which some drivers could have an
overcautious behavior, if all drivers behave in the same way,
then all cars will stop for a density less than the maximum
car density $\rho=1$.\footnote{In CA traffic
models, a cell can accomodate at most one car.} We will,
therefore, not consider such a driving strategy acceptable for
deterministic CA traffic rules.

\subsection{Local jams} The existence of a free-moving phase
implies the existence of a mechanism making possible the
annihilation of all local jams. For deterministic FI models, the
jammed car just behind the free-moving car leading the local jam
becomes free to move at the next time step. If, at low density,
we want local jams to gradually disappear, this condition,
which was automatically satisfied in the case of deterministic
FI traffic flow models, should be required for models evolving
according to unidirectional number-conserving deterministic CA
rules to be acceptable traffic rules.

\vspace{0.2cm}

\textbf{Definition 3} \textsl{A system of particles evolving
according to a unidirectional velocity rule is a deterministic
traffic flow model if a sequence of defective tiles cannot be
preceded by free empty cells. A sequence of defective tiles
preceded by a perfect tile is called a local jam.}

\vspace{0.2cm}

Proposition 1 can then be extended to all deterministic traffic
flow CA models as defined above.

\vspace{0.2cm}

\textbf{Proposition 4} \textsl{In a deterministic CA model of
traffic flow, the number of cars belonging to a local jam is a
non-increasing function of time.}

\vspace{0.2cm}

\textbf{Example 7} A system of particles evolving according
to the velocity rule:
\begin{equation}
\mathtt{\bullet\!\bullet\!11\bullet\rightarrow 0},\quad
\mathtt{\bullet 0101\rightarrow 0},\quad
\mathtt{\bullet 110\bullet\rightarrow 1},\quad
\mathtt{\bullet 0100\rightarrow 2}
\label{bad1}
\end{equation}
is not a traffic flow model. All configurations are
concatenations of the following three types
of tiles and free empty cells:
\begin{center}
\begin{tabular}{|c|c|c|}
\hline
\cell{\tt 2}&\cell{\tt e}&\cell{\tt e}\\
\hline
\end{tabular}
\hspace{0.7cm}
\begin{tabular}{|c|c|c|}
\hline
\cell{\tt 0}&\cell{\tt 1}&\cell{\tt e}\\
\hline
\end{tabular}
\hspace{0.7cm}
\begin{tabular}{|c|c|}
\hline
\cell{\tt 0}&\cell{\tt e}\\
\hline
\end{tabular}
\end{center}
The first tile is the perfect tile while the other two tiles are
the only defective tiles. Since to move at $v_{\max}=2$ a particle
needs not only to have two empty sites in front but also one empty
site behind, the existence of a defective tile which contains two
particles makes that, as shown in the example below, a sequence
of defective tiles might be preceded by free empty cells.
\begin{align*}
\mbox{time $t$:}
\quad & \cdots e e e e 2 e e 0 1 e 0 e 2 e e e e \cdots\\
\mbox{time $t+1$:}
\quad & \cdots e e e e e e 0 1 e 0 1 e e e 2 e e \cdots
\end{align*}
To avoid this behavior a defective tile should consist of a cell
occupied by one particle with velocity $v<v_{\max}$ preceded by $n_v$
empty cells, where $v\leq n_v<v_{\max}$.

For FI models, the variational principle followed from the
particular structure of defective tiles, that is, the number $n_v$
of empty cells preceding a cell in state $v$ was always equal to $v$.
All models in which defective tiles have this structure will,
therefore, verify Proposition 3, and we can state:

\vspace{0.2cm}

\textbf{Proposition 5} \textsl{If a deterministic CA traffic rule is
such that local jams have the following structure
$$
\cdots v_1\underbrace{ e e \cdots e}_{v_1}
v_2\underbrace{e e \cdots e}_{v_2}
\cdots\cdots v_n\underbrace{e e \cdots e}_{v_n}
v_{\max}\underbrace{e e \cdots e}_{v_{\max}}\cdots
$$
where, for all $i\in\{1,2,\ldots,n\}$, $0\leq v_i<v_{\max}$, then,
the average car flow $\rho\la v\ra_t$ is a non-decreasing function
of time $t$, and reaches its maximum value in the steady state.}

\vspace{0.2cm}

\begin{figure}[ht]
\begin{center}
\includegraphics[scale=1.2]{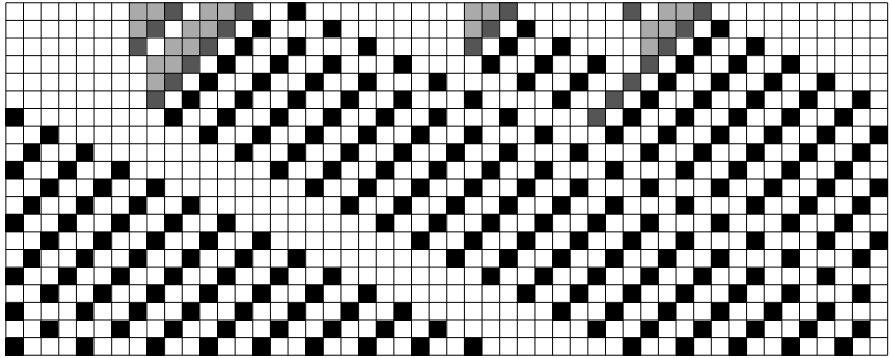}
\caption{\small{Spatiotemporal pattern of rule (\ref{R15}) for $\rho=0.28$.
States color code: white = $e$, light grey = $0$, dark grey = $1$,
black = 2.}}
\label{R15d028}
\end{center}
\end{figure}

\textbf{Example 8} The velocity rule
\begin{equation}
\mathtt{\bullet\bullet 11\bullet\rightarrow 0},\quad
\mathtt{\bullet 110\bullet\rightarrow 1},\quad
\mathtt{\bullet 0101\rightarrow 1},\quad
\mathtt{\bullet 0100\rightarrow 2}
\label{R15}
\end{equation}
is a nontrivial CA traffic rule satisfying Proposition 5. The
speed limit $v_{\max}$ is equal to $2$. As shown in
Figure \ref{R15d028}, at low density, all cars move at $v_{\max}$.
In this regime, the configurations belonging to the limit set consist
of perfect three-cell tiles of the type
\begin{center}
\begin{tabular}{|c|c|c|}
\hline
\cell{\tt 2}&\cell{\tt e}&\cell{\tt e}\\
\hline
\end{tabular}
\end{center}
in a sea of free empty cells. The critical density $\rho_c$ is
equal to $\frac{1}{3}$.

\begin{figure}[ht]
\begin{center}
\includegraphics[scale=1.2]{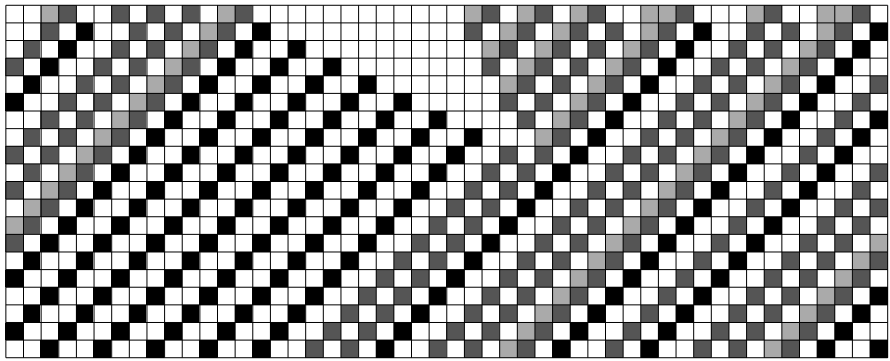}
\caption{\small{Spatiotemporal pattern of rule (\ref{R15}) for
$\rho=0.42$. States color code: white = $e$, light grey = $0$,
dark grey = $1$, black = $2$.}}
\label{R15d042}
\end{center}
\end{figure}

Figure \ref{R15d042} shows a spatiotemporal pattern for a car density
higher than the critical density.
All configurations, except the initial configuration chosen at random,
are concatenations of perfect tiles, defective tiles, of the following
two types:
\begin{center}
\begin{tabular}{|c|c|}
\hline
\cell{\tt 1}&\cell{\tt e}\\
\hline
\end{tabular}
\hspace{1cm}
\begin{tabular}{|c|}
\hline
\cell{\tt 0}\\
\hline
\end{tabular}
\end{center}
and free empty cells. Eventually all free empty cells disappear.
In both cases, except for the initial configuration, all local
jams are such that Proposition 5 applies.

\textbf{Example 9} There are many nontrivial traffic rules to
which Proposition 5 applies. The following traffic rule
represents another example:
\begin{align}
& \mathtt{\bullet\!\bullet\!\bullet 11\bullet\!\bullet
\rightarrow 0},\quad
\mathtt{\bullet\!\bullet\!\bullet\!101\bullet
\rightarrow 1},\quad
\mathtt{\bullet\!\bullet 01001 \rightarrow 2},\notag\\
& \qquad\quad \mathtt{\bullet\!\bullet\!1100\bullet \rightarrow 2},\quad
\mathtt{\bullet\!\bullet 01000\rightarrow 3}
\label{R1rad3}
\end{align}

\textbf{Remark 4} If, as in Example 3, we denote by $\rho_v$
the fraction of all particles whose velocities are equal to $v$
in the steady state, for all velocity rules to which
Proposition 5 applies, we have
\begin{align}
\sum_{v=0}^{v_{\max}} \rho_v & = \rho\label{rhov1}\\
\sum_{v=0}^{v_{\max}} (v+1) \rho_v & = 1.\label{rhov2}
\end{align}
Subtracting (\ref{rhov1}) from (\ref{rhov2}), we find that the
asymptotic average velocity is given by
$$
\la v\ra_\infty = \frac{1}{\rho}\sum_{v=0}^{v_{\max}} v \rho_v
= \frac{1-\rho}{\rho}.
$$
The last expression is exactly the result one obtains using
a mean-field argument. The fact that mean-field arguments lead to
exact results for CA traffic rules to which Proposition 5 applies
has already been found for the values of the critical exponents of
phase transitions in FI traffic flow models \cite{BF3}. It might
be conjectured that, for all CA traffic rules to which
Proposition 5 applies, critical exponents will be equal to
their mean-field values.

\textbf{Example 10} Proposition 5 does not apply to a system of
particles evolving according to the rule:
\begin{align}
&\mathtt{\bullet\!\bullet\!11\bullet\rightarrow 0},\quad
\mathtt{\bullet\!\bullet\!101\rightarrow 0},\quad
\mathtt{\bullet\!\bullet\!100\rightarrow 2}
\label{Rcr64}
\end{align}
since all jammed cars have a zero velocity even when there exists
a free empty cell in front. All configurations are concatenations
of the following three types of tiles:
\begin{center}
\begin{tabular}{|c|c|c|}
\hline
\cell{\tt 2}&\cell{\tt e}&\cell{\tt e}\\
\hline
\end{tabular}
\hspace{0.7cm}
\begin{tabular}{|c|c|}
\hline
\cell{\tt 0}&\cell{\tt e}\\
\hline
\end{tabular}
\hspace{0.7cm}
\begin{tabular}{|c|}
\hline
\cell{\tt 0}\\
\hline
\end{tabular}
\end{center}
and free empty cells. The first tile, which corresponds to cars
moving at the speed limit $v_{\max}=2$, is the perfect tile, and
the other two tiles, corresponding to cars with a zero velocity
are the only defective tiles. It is easy to verify, however, that
Proposition 4 apply. Since all jammed cars are stopped cars, it
follows that, for a given car density, the average velocity
$\la v\ra_t$ is a monotonous non-increasing function of time.

\textbf{Example 11} For systems evolving according to the rule:
\begin{align}
& \mathtt{\bullet\!\bullet\!\bullet 11\bullet\!\bullet
\rightarrow 0},\quad
\mathtt{\bullet\!\bullet\!\bullet\!101\bullet
\rightarrow 0},\quad
\mathtt{\bullet\!\bullet 01001 \rightarrow 1},\notag\\
& \qquad\quad \mathtt{\bullet\!\bullet\!11001 \rightarrow 2},\quad
\mathtt{\bullet\!\bullet\!\bullet 1000\rightarrow 3},
\label{R2rad3}
\end{align}
due to the existence of defective tiles of the following structures:
\begin{center}
\begin{tabular}{|c|c|c|}
\hline
\cell{\tt 1}&\cell{\tt e}&\cell{\tt e}\\
\hline
\end{tabular}
\hspace{1cm}
\begin{tabular}{|c|c|}
\hline
\cell{\tt 0}&\cell{\tt e}\\
\hline
\end{tabular}
\end{center}
Proposition 5 does not apply, and while Proposition 4 applies,
for a given car density, the average velocity $\la v\ra_t$
is not a monotonous non-increasing functions of time, since,
as shown below, the velocity of a jammed car, equal to 1 at
time $t$, may become equal to 0 at time $t+1$:
\begin{align*}
\mbox{time $t$:}
\quad & \cdots 3 e e e 1 e e 3 e e e e e e \cdots\\
\mbox{time $t+1$:}
\quad &  \ \ \cdots e e 0 e 3 e e e e 3 e e e \cdots.
\end{align*}

This example shows that, while a deterministic CA rule may be
an acceptable deterministic CA traffic rule, the
variational principle, which in its stronger form states that,
\textit{for a given car density, the average flow is a monotonous
non-decreasing function of time}, does not apply. The question
which could be addressed is then: Is, however, the variational
principle still valid under a weaker form?

Since we assumed that a system evolving according to a
deterministic CA rule should exhibit a free-moving phase at low
car density, by definition of the free-moving phase, all cars
move at $v_{\max}$. Therefore, for $\rho<\rho_c$,
\textit{the average car flow takes its maximum value in the
steady state}. Above the critical density, it is only if the
asymptotic average flow is a well-defined function of car density
that the question makes sense. If this is the case, then, since
Proposition 4 applies to all traffic flow deterministic CA models,
it may be reasonably conjectured that, also for $\rho>\rho_c$, the
average car flow takes its maximum value in the steady state.

\textbf{Remark 5} This paper deals with the existence of a
variational principle for number-conserving deterministic CA
models of highway traffic flow. The variational principle, as
stated in Proposition 5, may be valid, however, for
one-dimensional closed systems of particles evolving according
to more general number-conserving deterministic CA rules. For
instance, in the case of Rule (\ref{R4}), presented Example 2, all
configurations are concatenations of the following three tiles:
\begin{center}
\begin{tabular}{|c|c|c|c|}
\hline
\cell{\tt 0}&\cell{\tt 2}&\cell{\tt e}&\cell{\tt e}\\
\hline
\end{tabular}
\hspace{1cm}
\begin{tabular}{|c|c|}
\hline
\cell{\tt 1}&\cell{\tt e}\\
\hline
\end{tabular}
\hspace{1cm}
\begin{tabular}{|c|}
\hline
\cell{\tt 0}\\
\hline
\end{tabular}
\end{center}
and free empty cells. The tiles' structure shows that Relation
(\ref{fec}) is valid. If the particle density $\rho$ is less than
$\frac{1}{3}$, there are enough free empty cells to annihilate all
cells of the first and third type; and all configurations of the
limit set are concatenations of tiles of the second type and free
empty cells. If $\rho>\frac{1}{3}$, there are not enough free
empty cells to annihilate all cells of the first and third type;
and all configurations of the limit set contain tiles of all types
but no free empty cells. Although such a system does
not exhibit a free-moving phase according to Definition 2, cells
of the second type play the role of ``perfect tiles'' while the
other two types of cells can be regarded as ``defective tiles''.

When Relation (\ref{fec}) is satisfied, it is necessary to verify
that the density of free empty cells is a non-increasing function
of time to ensure the validity of the variational principle under
its stronger form. For instance, while $\rho_{\rm fec}(t)$ cannot
increase with time $t$ for a system of particles evolving
according to Rule (\ref{R4}), this is no more the case
for a system of particles evolving according Rule (\ref{R8}),
presented in Example 5, and which is obtained by replacing
the element $\mathtt{\bullet 010\bullet\rightarrow 1}$ of Rule
(\ref{R4}) by the two elements $\mathtt{1010\bullet\rightarrow 0}$
and $\mathtt{0010\bullet \rightarrow 1}$.
As for Rule (\ref{R4}), all configurations are concatenations
of the same above three tiles and free empty cells, which implies
that Relation (\ref{fec}) is satisfied. But, due the
structure of the element $\mathtt{1010\bullet\rightarrow 0}$
of Rule (\ref{R8}), as shown below (three-particle system
evolving on a size-10 lattice):
\begin{align*}
\mbox{time $t$}\quad & \underline{e e} 0 2 e e 1 e
\underline{e e}\\
\mbox{time $t+1$}\quad & \underline{e e} 1 e
\underline{e} 1 e 0 \underline{e e}
\end{align*}
the number of free empty cells ($\underline{e}$) increases
from 4 at time $t$ to 5 at time $t+1$ resulting in a decrease
of the average velocity.

\section{Conclusion} As illustrated by a number of different examples,
many number-conserving deterministic CA rules cannot be considered
reasonable deterministic traffic rules, and we tried to list the
essential properties characterizing a traffic rule in order to give
a general definition of deterministic CA models of highway traffic flow.
Various notions such as free-moving phase, perfect and defective tiles,
and local jam play an important role in our discussion. We have then
shown that, within the framework of this definition, a variety of
deterministic CA models of traffic flow obey a variational principle
which, in its stronger form, states that, for a given car density,
the average car flow is a non-decreasing function of time. This
result has been established for traffic flow models whose
configurations exhibits local jams of a given structure. If local
jams have a different structure, while this variational principle
may still apply to systems evolving according to some particular
rules, it will not apply in general. However, if the asymptotic
average car flow is a well-defined function of car density,
since we have proved that for all traffic flow models
the number of jammed cars of a local jam cannot increase, we
conjectured that it will apply under a weaker form which states
that, for a given car density, the average car flow takes its
maximum value in the steady state.

The variational principle also applies, even in its stronger
form, to many number-conserving deterministic CA rules which
cannot be considered reasonable traffic rules.

\vspace{1cm}

\noindent\textbf{Acknowledgements}

\vspace{0.3cm}

The author is grateful to Henryk Fuk\'s and Andr\'es Moreira for
their very good suggestions. This work has been done during a stay
at the Centro de Modelamiento Matem\'atico de la Universidad
de Chile in Santiago thanks to FONDAP-CONICYT. The unflagging
interest of Eric Goles has been of great help.


\newpage

\begin{thebibliography}{0000}

\bibitem{NaSch} K. Nagel and M. Schreckenberg,
J. Physique (France) {\bf I2} 2221 (1992).

\bibitem{TT} M. Takayasu and H. Takayasu, Fractals {\bf 1} 860 (1993)

\bibitem{FI} M. Fukui and Y. Ishibashi, J. Phys. Soc. Japan
{\bf 65}, 1868 (1996)

\bibitem{TTMS} T. Tokihiro, D. Takahashi, J. Matsukidaira and J.
Satsuma, Phys, Rev. Lett. {\bf 76} 3247 (1996)

\bibitem{FB} H. Fuk{\'s} and N. Boccara, Int. J. Mod. Phys. C
{\bf 9} 1 (1998),\\
(H. Fuk\'s and N. Boccara 1997 \textit{Preprint}
\texttt{adap-org/9705003}).

\bibitem{NT1} K. Nishinari and D. Takahashi, J. Phys. A: Math.
Gen. {\bf 32} 93 (1999)

\bibitem{W} S. Wolfram S, {\it Cellular Automata and Complexity:
Collected Papers} (Reading, Massachusetts: Addison-Wesley, 1994).

\bibitem{F} H. Fuk{\'s}, Phys. Rev. E {\bf 60}, 197 (1999),\\
(H. Fuk\'s 1999 \textit{Preprint} \texttt{comp-gas/9902001}).

\bibitem{BF1} N. Boccara and H. Fuk\'s, to appear in Fundamenta
Informaticae,\\
(N. Boccara and H. Fuk\'s 1997 \textit{Preprint}
\texttt{cond-mat/9905004}).

\bibitem{NT2} (K. Nishinari and D. Takahashi 2000
\textit{Preprint} \texttt{nlin.AO/0002007}

\bibitem{BF2} N. Boccara and H. Fuk{\'s}, J. Phys. A: Math. Gen.,
{\bf 31} 6007 (1998),\\
(N. Boccara and H. Fuk\'s 1997 \textit{Preprint}
\texttt{adap-org/9712003}).

\bibitem{Schad} (A. Schadschneider 1999 \textit{Preprint}
\texttt{cond-mat/9902170}).

\bibitem{BF3} N. Boccara and H. Fuk{\'s}, to appear in volume
{\bf 33} (2000) of the J. of Phys. A: Math. Gen.,\\
(N. Boccara and H. Fuk\'s 1999 \textit{Preprint}
\texttt{cond-mat/9911039}).

\end{thebibliography}
\end{document}